# Sign-free stochastic mean-field approach to strongly correlated phases of ultracold fermions.

**O JUILLET**

LPC/ENSICAEN, Boulevard du Maréchal Juin, F-14050 Caen Cedex, France

E-mail: juillet@lpccaen.in2p3.fr

**Abstract.** We propose a new projector quantum Monte-Carlo method to investigate the ground state of ultracold fermionic atoms modeled by a lattice Hamiltonian with on-site interaction. The many-body state is reconstructed from Slater determinants that randomly evolve in imaginary-time according to a stochastic mean-field motion. The dynamics prohibits the crossing of the exact nodal surface and no sign problem occurs in the Monte-Carlo estimate of observables. The method is applied to calculate ground-state energies and correlation functions of the repulsive two-dimensional Hubbard model. Numerical results for the unitary Fermi gas validate simulations with nodal constraints.

**PACS.** 03.75.Ss, 05.30.Fk, 71.10.Fd

**1. Introduction.** Since the experimental achievement of Fermi degeneracy [1] with an atomic vapor, a considerable attention has been attracted by the physics of dilute ultracold fermions. The ability to tune many parameters, such as temperature, density or inter-particle interactions, makes atomic Fermi gases ideal candidates to understand a wealth of phenomena relevant for physical systems ranging from nuclear matter to high-temperature superconductors. Of particular interest is the strongly interacting regime in the transition from Bardeen-Cooper-Schrieffer (BCS) superfluidity of Cooper pairs to Bose-Einstein condensation (BEC) of atomic dimers. In the crossover regime, Fermi condensates have been observed on both the BCS and the BEC sides of a magnetically controlled Feshbach resonance [2-4]. The unitary regime, where the scattering length diverges, is particularly studied [5-7] to investigate the universal features of the fermionic quantum many-body problem. Using several standing laser beams, ultracold atoms can also be loaded in optical lattices where they experience all the strong many-body correlations described by the Hubbard model of solid-state physics [8,9]. Optical lattice setups may allow for engineering quantum spin models [10], fractional quantum Hall effect [11], non-Abelian gauge potentials [12] or quantum information processing [13].
In this paper, we investigate a new Monte-Carlo scheme to study strongly correlated ground states of ultracold fermions interacting on a lattice. The projection onto the ground state is performed through a reformulation of the imaginary-time Schrödinger equation in terms of Slater determinants undergoing a Brownian motion driven by the Hartree-Fock Hamiltonian. Such exact stochastic extensions of the mean-field approaches have been recently proposed for boson systems [14,15]. Up to now, the fermionic counterpart uses Slater determinants whose orbitals evolve under their own mean-field, supplemented with a stochastic one-particle-one-hole excitation [16,17]. Unfortunately, the sampling generally suffers from negative weight trajectories that cause an exponential decay of the signal-to-noise ratio, which is known as the sign problem. The convergence issue of such a Monte-Carlo calculation plagued by negative "probabilities" belongs to the class of NP hard problems and a polynomial complexity solution can be probably ruled out [18]. Here, we extend the stochastic Hartree-Fock approach to remove negative weight paths in the Monte-Carlo calculation of any observable.

**2. The sign-free stochastic mean-field scheme.** For a system of fermions interacting through a binary potential, we first introduce a set of hermitian one-body operators $\hat{A}_s$ $(s = 0,1,\cdots)$ allowing to rewrite the model Hamiltonian $\hat{H}$ as a quadratic form:

$$\hat{H} = \hat{A}_0 - \sum_{s \geq 1} \omega_s \hat{A}_s^2 \ , \quad \hat{A}_s = \sum_{i,j} (A_s)_{ij} \hat{a}_i^+ \hat{a}_j \tag{1}$$

$\hat{a}_i^+, \hat{a}_i$ are the Fermi creation and annihilation operator in a single-particle mode $|i\rangle$. Let us now consider two $N$-particle Slater determinants $|\psi\rangle, |\varphi\rangle$ of orbitals satisfying the biorthogonality relations $\langle \psi_n | \varphi_p \rangle = \delta_{n,p}$. Any matrix element $\langle \psi | \hat{A} | \varphi \rangle$ can then evaluated by using Wick's theorem with the set of contractions $\mathcal{R}_{ij} = \langle \psi | \hat{a}_j^+ \hat{a}_i | \varphi \rangle = \sum_n \langle i | \varphi_n \rangle \langle \psi_n | j \rangle$. The non-hermitian one-body density matrix $\mathcal{R}$ is linked to the usual Hartree-Fock single-particle Hamiltonian $h$ by $h_{ij}(\mathcal{R}) = \frac{\partial \mathcal{H}}{\partial \mathcal{R}_{ji}}$, where $\mathcal{H}(\mathcal{R}) = \langle \psi | \hat{H} | \varphi \rangle$ is the matrix element of the Hamiltonian. Under the Hamiltonian (1), it is shown in Appendix that the Slater determinant $|\varphi\rangle$ transforms according to:

$$\hat{H}|\varphi\rangle = \mathcal{H}|\varphi\rangle + \hat{h}_{0,1}|\varphi\rangle - \sum_{s \geq 1} \omega_s \hat{A}_{s,0,1}^2 |\varphi\rangle \tag{2}$$

$\hat{h}_{0,1}$ is defined by $\hat{h}_{0,1} = \sum_{i,j} [(1-\mathcal{R})h\mathcal{R}]_{ij} \hat{a}_i^+ \hat{a}_j$ and $\hat{A}_{s,0,1}$ is given by similar expression: $\hat{A}_{s,0,1} = \sum_{i,j} [(1-\mathcal{R})A_s \mathcal{R}]_{ij} \hat{a}_i^+ \hat{a}_j$. If the two Slater determinants $|\psi\rangle, |\varphi\rangle$ were identical, $\hat{h}_{0,1}$ and $\hat{A}_{s,0,1}^2$ would represent physically one-particle-one-hole and two-particle-two-holes excitations. After an infinitesimal time step $d\tau$, during which $|\varphi\rangle$ evolves to $\exp(-d\tau \hat{H})|\varphi\rangle$, the last term of the expansion (2) causes departures of the propagated state from a single Slater determinant. However, introducing random fields according to the Hubbard-Stratonovich transformation can linearize the dynamics [19-21]. The detailed derivation is reported in Appendix. Finally, the exact imaginary-time evolution is recovered through the stochastic average of Brownian trajectories in the subspace of Slater determinants states $|\varphi\rangle$:

$$\exp(-\tau \hat{H})|\varphi(0)\rangle = E[\exp(S(\tau)) |\varphi(\tau)\rangle] \tag{3}$$

where $E(\cdots)$ is the average over a random functional; the evolution of $S$ and the orbitals $|\varphi_n\rangle$ is governed by the following equations in the Itô sense:

$$dS = -d\tau \, \mathcal{H}(\mathcal{R}), \quad S(0) = 0 \tag{4}$$

$$d|\varphi_n\rangle = (1-\mathcal{R})\left(-d\tau \, h(\mathcal{R}) + \sum_{s \geq 1} dW_s \sqrt{2\omega_s} A_s\right)|\varphi_n\rangle. \tag{5}$$

The $dW_s$ are infinitesimal increments of independent Wiener processes: $E(dW_s) = 0$, $dW_s dW_{s'} = d\tau \, \delta_{s,s'}$. We emphasize that the dynamics (5) exactly preserves the biorthogonality properties between the two Slater determinants $|\psi\rangle, |\varphi\rangle$: indeed, the left-eigenvalue equation $\langle \psi_n | \mathcal{R} = \langle \psi_n |$ implies $\langle \psi_n | \varphi_p + d\varphi_p \rangle = \delta_{np}$ and therefore $\langle \psi | \varphi \rangle = 1$ at any time. This feature guarantees that sign problems will not occur as long as $S$ remains real during the imaginary-time motion, as we detail below.

Consider a walker $|\varphi(\tau_o)\rangle$ at time $\tau_o$. Its overlap with the exact many-body ground-state $|\Psi_g\rangle$ can be obtained from the representation (3) in the limit of large $\tau$, provided that the trial state $|\psi\rangle$ is not orthogonal to $|\Psi_g\rangle$:

$$\langle \Psi_g | \varphi(\tau_o) \rangle = \lim_{\tau \to \infty} \frac{\langle \psi | \exp(-\tau \hat{H}) | \varphi(\tau_o) \rangle}{\langle \psi | \Psi_g \rangle \exp(-\tau E_g)} = \lim_{\tau \to \infty} \frac{E[\exp(S(\tau))\langle \psi | \varphi(\tau) \rangle]}{\langle \psi | \Psi_g \rangle \exp(-\tau E_g)} \quad (6)$$

where $E_g$ is the ground-state energy and where the phase of $|\psi\rangle$ can be chosen to have $\langle \psi | \Psi_g \rangle$ real and positive. With the dynamics (5) of the Hartree-Fock orbitals, $\langle \psi | \varphi \rangle = 1$ and no walker can cross the exact nodal surface if $S$ is real. In contrast, the standard auxiliary-field projector quantum Monte-Carlo method [22], as well as our previous stochastic mean-field scheme [16,17], generally leads to Slater determinants $|\varphi\rangle$ whose overlap $\langle \psi | \varphi \rangle$ exhibits a varying sign. As a consequence, a walker can reach the nodal surface and it then generates stochastic paths that do not contribute to the ground-state: $\langle \Psi_g | \exp(-\tau \hat{H}) | \varphi(\tau_o) \rangle = E[\exp(S(\tau)\langle \Psi_g | \varphi(\tau) \rangle] = 0$ as long as $\langle \Psi_g | \varphi(\tau_o) \rangle = 0$. Such trajectories only increase the statistical error and are responsible of the sign-problem. Thus, one is forced to perform the constrained-path approximation [23] where walkers are eliminated "by hand" when their overlap with a ground-state ansatz wave-function becomes negative.

Numerically, the stochastic differential equations (5) are solved, in the Stratonovich form, by an embedded Runge-Kutta (5,4) algorithm with adaptive stepsize control. We take $|\varphi(0)\rangle = |\psi\rangle$ as initial condition and the spreading of the weights $\exp(S)$ is avoided through standard population control techniques. In practice, we use the stochastic reconfiguration method [24] that deals with a fixed-number of walkers $|\varphi\rangle$ among which some are killed and others are cloned according to their relative weight in the population. Observables are estimated from the representation (3) of the many-body state. For instance, the ground-state energy $E_g$ is be obtained according to

$$E_g = \lim_{\tau \to \infty} \frac{\langle \psi | \hat{H} \exp(-\tau \hat{H}) | \varphi(0) \rangle}{\langle \psi | \exp(-\tau \hat{H}) | \varphi(0) \rangle} = \lim_{\tau \to \infty} \frac{E[\exp(S(\tau))\langle \psi | \hat{H} | \varphi(\tau) \rangle]}{E[\exp(S(\tau))]} \quad (7)$$

where the amplitudes $\exp(S(\tau))$ are now all real positive. Moreover, when the Slater determinant ansatz $|\psi\rangle$ and the ground-state share a common symmetry, the stochastic paths are automatically projected onto this symmetry sector in the estimate (7). Otherwise, the sampling can be improved by projection techniques [25,26]. These conclusions also hold true for any observable commuting with the Hamiltonian. In other cases, one obtains an approximate ground-state expectation value, known as the mixed estimate [27]. It can be corrected by the following extrapolated estimate that is one order better in the difference $|\psi\rangle - |\Psi_g\rangle$ [27]:

$$\langle \hat{O} \rangle_{extrap.} = 2\langle \hat{O} \rangle_{mixed.} - \langle \psi | \hat{O} | \psi \rangle, \quad \langle \hat{O} \rangle_{mixed.} = \lim_{\tau \to \infty} \frac{E\left[\exp(S(\tau)) \langle \psi | \hat{O} | \varphi(\tau) \rangle \right]}{E[\exp(S(\tau))]} \qquad (8)$$

Note that these observable estimates are biased in standard projector quantum Monte-Carlo methods by the nodal constraints introduced to circumvent sign problems. The stochastic Hartree-Fock approach (3-5) removes this systematic error. The exact expectation value would require obtaining the ground-state density matrix by a double propagation in imaginary-time: $|\Psi_g\rangle\langle\Psi_g| \propto \lim_{\tau \to \infty} \exp\left(-\frac{\tau}{2}\hat{H}\right)|\psi\rangle\langle\psi|\exp\left(-\frac{\tau}{2}\hat{H}\right)$. For such calculations, our scheme can be extended by expanding the density-matrix in terms of dyadics $|\varphi^{(a)}\rangle\langle\varphi^{(b)}|$ that are formed by biorthogonal Slater determinants, both undergoing a Brownian motion similar to Eq. (5). We emphasize that the method then becomes the equivalent, at fixed particle-number, of the recent Gaussian Monte-Carlo technique [25,28], which is more suited for thermodynamical studies in the grand-canonical ensemble.

**3. Application to ultracold atomic Fermi gases.** First, we concentrate on the single-band $SU(2)$ Hubbard model that describes the low-energy physics of two-component ultracold fermions trapped in optical lattices:

$$\hat{H} = -t \sum_{\langle \vec{r}, \vec{r}' \rangle, \sigma=\uparrow,\downarrow} \hat{a}^+_{\vec{r},\sigma} \hat{a}_{\vec{r}',\sigma} + U \sum_{\vec{r}} \hat{n}_{\vec{r},\uparrow} \hat{n}_{\vec{r},\downarrow} \qquad (9)$$

Here $\hat{a}^+_{\vec{r},\sigma}$ creates one atom at site $\vec{r}$ in the internal state $|\sigma\rangle = |\uparrow\rangle, |\downarrow\rangle$ and $\hat{n}_{\vec{r},\sigma} = \hat{a}^+_{\vec{r},\sigma} \hat{a}_{\vec{r},\sigma}$ is the corresponding number operator; $t$ is the hopping matrix element between nearest neighboring sites $\langle \vec{r}, \vec{r}' \rangle$; $U$ is the amplitude of the on-site interaction between two atoms. Analytical solutions only exist in one dimension. For higher dimensional problems, standard auxilliary-field quantum Monte-Carlo calculations are limited to the repulsive model at half filling and to the attractive model with symmetric populations in the two spin channels $|\uparrow\rangle, |\downarrow\rangle$. In other cases, one experiences severe sign problems that practically prohibit studying large lattices, strongly interacting systems or open shells configurations [22]. In contrast, our new stochastic Hartree-Fock scheme (3-5) does not manifest explicit sign problems regardless of the lattice topology, band filling and sign of the interaction. Indeed, a quadratic form (1) can be recovered from the Hamiltonian (9) by using the local density or magnetization depending on the sign of the interaction parameter $U$:

$$\hat{H} = -t \sum_{\langle \vec{r}, \vec{r}' \rangle, \sigma=\uparrow,\downarrow} \hat{a}^+_{\vec{r},\sigma} \hat{a}_{\vec{r}',\sigma} - \frac{|U|}{2} \sum_{\vec{r}} \left( \hat{n}_{\vec{r},\uparrow} + \mathrm{sgn}(U) \hat{n}_{\vec{r},\downarrow} \right)^2 \qquad (10)$$

where we have omitted a constant term proportional to the total number of particles. All the one-body operators $\hat{A}_s$ defined by Eq. (10) are real in the representation $|\vec{r},\sigma\rangle$. For real orbitals of the trial Slater determinant $|\psi\rangle$ in this basis, their biorthogonal partners $|\varphi_n\rangle$, stochastically propagated by the dynamics (5), are real at any imaginary-time as well. Therefore, $\mathcal{H}(\mathcal{R})$ and $S(\tau)$ are also real, and positive weights trajectories are guaranteed. For the positive-$U$ model (9), our ansatz $|\psi\rangle$ is a spin-singlet Slater determinant for free-fermions $|\psi_F\rangle$ or an antiferromagnetic Hartree-Fock mean-field solution $|\psi_{AF}\rangle$.

As an example, we consider an unpolarized system of 12 atoms interacting on a $4\times 4$ lattice: if $U=4t$, the exact ground-state energy is $E_g=-17.73\,t$ [29] and the stochastic Hartree-Fock propagation of $|\psi_F\rangle$ (resp. $|\psi_{AF}\rangle$) leads to the value $-17.709(7)$ (resp. $-17.710(6)$) for the energy estimator (7) at the imaginary-time $\tau=20/t$. Other results are displayed on Fig. 1 for two-dimensional lattices with a hole doping $\delta=0.125$ from half-filling. This density generates the most important sign problem in auxiliary-field approaches [22]. The antiferromagnetic mean-field state is very efficient and the exact ground-state energy of the $4\times 4$ lattice at $U=4t$ is recovered to within less than 0.2% if the sampling is improved by projecting onto the spin-singlet sector. For the larger on-site repulsion $U=8t$, the stochastic Hartree-Fock estimate of the ground-state energy is $E_g=-11.845(1)\,t$ ($\tau=20/t$) as compared to the exact value $E_g=-11.872\,t$. The simulations on the $8\times 8$ lattice, presented on the right panel of Fig. 1, are in agreement to less than 0.5% with the constrained-path approach [23]. This is fully consistent with the error usually observed in constrained-path calculations of the ground-state energy on small clusters [23]. But, we emphasize that such small discrepancies can originate from the numerical error in the integration of the stochastic differential equations (5) and from the unavoidable bias that is introduced by the population control algorithm.

On contrary, the finite-time simulations with the free-atom trial wave-function $|\psi_F\rangle$ does not achieve to filter out all the excited states near half-filling. Indeed, $|\psi_F\rangle$ is quasi-orthogonal to the ground-state $|\Psi_g\rangle$: for the $4\times 4$ cluster with $U=4t$, we have estimated the overlap $|\langle\psi_F|\Psi_g\rangle|^2$ to be around 0.0008 by approximating the ground-state vector to the stochastic mean-field results with the ansatz $|\psi_{AF}\rangle$ at $\tau=5/t$. As a consequence, the imaginary-time projection drives walkers into directions almost orthogonal to $|\psi_F\rangle$. This can be illustrated through the angle $\theta$ between a walker $|\varphi\rangle$ and the trial state $|\psi_F\rangle$: $\cos\theta=\langle\psi_F|\varphi\rangle/\|\varphi\|=1/\|\varphi\|$ and one obtains $E[\cos\theta]=0.0228$ for the time $\tau=5/t$. Unfortunately, such relevant trajectories with $\cos\theta\approx 0$ can only appear when the norm $\|\varphi\|$ reaches large values, leading to numerical errors both in the dynamical evolution of the wave-function and in the calculation of observables. The Monte-Carlo sampling can also become incorrect if a power-law tail develops in the walkers' distribution. This scenario would then be identical to the one encountered in real-time simulations of boson systems in the context of positive-P representation [30].

In the half-filling limit, our calculations confirm the emergence of an antiferromagnetic phase [31], as shown in Fig. 2 through the extrapolated estimate (8) of the space spin-spin correlation function $\langle \hat{\vec{S}}_{\vec{0}}\cdot\hat{\vec{S}}_{\vec{r}}\rangle$. The antiferromagnetic order is destroyed by hole doping or by a geometrical frustration induced via a large next-nearest neighbor hopping (see Fig.2), in agreement with Ref. [32]. The extrapolated values (8) of the magnetic and charge structure factors, defined by

$$S_m(\vec{q})=\frac{4}{3}\sum_{\vec{r}} e^{i\vec{q}\cdot\vec{r}}\langle \hat{\vec{S}}_{\vec{0}}\cdot\hat{\vec{S}}_{\vec{r}}\rangle,\ S_c(\vec{q})=\sum_{\vec{r}} e^{i\vec{q}\cdot\vec{r}}\langle \hat{n}_{\vec{0}}\hat{n}_{\vec{r}}\rangle, \qquad (11)$$

are consistent with the diagonalization results on the $4\times 4$ lattice [33]: at the corner $\vec{q}=(\pi,\pi)$ of the Brillouin zone, one obtains $S_m=3.681(4), S_c=0.3872(6)$ and $S_m=2.254(1), S_c=0.4236(1)$ compared to the exact values $S_m=3.64, S_c=0.385$ and $S_m=2.18, S_c=0.424$ for a doping $\delta=0$ (half-filling) and $\delta=0.125$, respectively.

We finally address the unitary Fermi gas limit. In this ideal regime of strong interaction via a two-body potential of zero range and infinite scattering length, fermions are among the most intriguing physical systems since they are believed to exhibit universal many-body states. For instance, at zero temperature, the energy must be a universal fraction $\eta$ of the Fermi energy that is the only relevant energy scale in the system. From experiments with trapped atomic gases [5-7], the measured values for this ratio $\eta$ vary from $0.32^{+10}_{-13}$ to $0.51(4)$. We model a spatially homogeneous Fermi gas by a lattice Hamiltonian with a two-body discrete delta potential whose coupling constant is adjusted to reproduce the physical scattering length $a$ [34]:

$$\hat{H} = \sum_{\vec{r},\vec{r}',\sigma=\uparrow,\downarrow} T_{\vec{r},\vec{r}'}\, \hat{a}^{+}_{\vec{r},\sigma} \hat{a}_{\vec{r}',\sigma} + \frac{4\pi\hbar^2}{Ml^3} \frac{a}{1-Ka/l} \sum_{\vec{r}} \hat{n}_{\vec{r},\uparrow} \hat{n}_{\vec{r},\downarrow} \qquad (12)$$

Here periodic boundary conditions are assumed in each direction; $T_{\vec{r},\vec{r}'}$ are the matrix elements of the single-particle kinetic energy operator in the representation position; $M$ is the atomic mass, $l$ denotes the grid spacing and $K = 2.44275\ldots$ is a numerical constant. In the unitary limit, $|a|$ goes to infinity but the coupling constant on the lattice remains finite and negative, so that the gas clearly experiences attraction. Our sign-free simulation method with the model Hamiltonian (12), transformed as in Eq. (10), has been checked from the known solutions of the two- and three-body problem in an harmonic trap at the unitarity point [35,36]: in all cases, the discrepancy does not exceed one percent. For different systems, up to $N = 42$ atoms on a $8\times 8\times 8$ lattice, we plot in Fig. 3 the convergence of the ratio $\eta(\tau)$ between the mean energy $\langle \hat{H} \rangle(\tau)$ of the unitary gas at the imaginary-time $\tau$ and the non-interacting ground-state energy $E_{g,0}$ on the lattice. In the limit of large $\tau$, all the results essentially concentrate around the same value and the emergence of a universal regime thus clearly appears. Fitting $\langle \hat{H} \rangle(\tau)$ by $E_{g,\infty} + \kappa e^{-\omega \tau}$, we estimate the ground-state energy $E_{g,\infty}$ at unitarity and find, for even particle-number, $\eta = E_{g,\infty}/E_{g,0} \approx 0.449(9)$ from the numerical values of Table 1. This result is consistent with recent exact Monte-Carlo calculations at finite-temperature [37,38]. It also validates fixed-node approaches that predict $\eta \approx 0.44(1)$ in the region $10 \leq N \leq 40$ and $\eta \approx 0.42(1)$ for larger systems [39,40]. For odd particle-number, we obtain similarly $E_{g,\infty} = 0.44(4) E_{g,0} + 0.442(3) \varepsilon_F$ where $\varepsilon_F$ is the Fermi level. Therefore, the empirical gap $\delta(N) = E_{g,\infty}(N) - (E_{g,\infty}(N-1) + E_{g,\infty}(N+1))/2$ displays the odd-even staggering characteristic of a superfluid. The odd-$N$ value of $\delta$, i.e. $0.442(3) \varepsilon_F$, gives an estimate of the pairing gap that is also of the same order as the fixed-node result $(0.54\, \varepsilon_F)$.

**4. Conclusion.** In summary, we have introduced a new stochastic Hartree-Fock scheme that allows quantum Monte-Carlo ground-state calculations of interacting fermions. For a wide class of ultracold fermions models, including the repulsive Hubbard model, positive weights trajectories are guaranteed and the sampling does not exhibit explicit sign problems. The method is in principle exact for the ground-state energy and for the mixed estimate of any observable. However, systematic errors can occur when the trial Slater determinant, that drives the stochastic motion, is closed to the nodal surface. Otherwise, the numerical simulations are very encouraging and accurate results have been obtained in situations that traditionally experience severe convergence problems. Further investigation on unbalanced resonant Fermi gases and doped Mott insulators are under development to provide new insights into the physics of strongly correlated fermions.

We acknowledge fruitful discussions with Y. Castin, R. Frésard and F. Gulminelli.

**Appendix. Derivation of the sign-free stochastic Hartree-Fock equations.**

Consider, at a given imaginary-time (omitted for simplicity), two $N$-particle Slater determinants $|\psi\rangle, |\varphi\rangle$ with biorthogonal orbitals:.

$$\langle \psi_n | \varphi_p \rangle = \delta_{n,p} \tag{A.1}$$

In a single-particle orthonormal basis $\{|i\rangle\}$, the generalized one-body density matrix $\mathcal{R}_{ij} = \langle \psi | \hat{a}_j^+ \hat{a}_i | \varphi \rangle = \sum_n \langle i | \varphi_n \rangle \langle \psi_n | j \rangle$ can always be transformed into the Jordan canonical form:

$$\Omega^{-1} \mathcal{R} \Omega = diag(J_1, J_2, \cdots, J_k, \cdots) \tag{A.2}$$

where $\Omega$ is a non-singular matrix and $J_k$ a $d_k \times d_k$ bi-diagonal matrix:

$$J_k = \begin{pmatrix} \lambda_k & 1 & 0 & \cdots & 0 & 0 \\ 0 & \lambda_k & 1 & \cdots & 0 & 0 \\ 0 & 0 & \lambda_k & \ddots & 0 & 0 \\ \vdots & \vdots & \vdots & \ddots & 1 & 0 \\ 0 & 0 & 0 & \cdots & \lambda_k & 1 \\ 0 & 0 & 0 & \cdots & 0 & \lambda_k \end{pmatrix} \tag{A.3}$$

Taking into account the idempotency of $\mathcal{R}$, i.e. $\mathcal{R}^2 = \mathcal{R}$, forces all the Jordan blocks $J_k$ to be one-dimensional $(d_k = 1)$ with $\lambda_k = 0, 1$. So, the column vectors of $\Omega$ and the line vectors of $\Omega^{-1}$ respectively correspond to the right eigenvectors $|\omega_k\rangle$ and to the left eigenvectors $\langle \widetilde{\omega}_k |$ of the non-hermitian diagonalizable matrix $\mathcal{R}$:

$$\mathcal{R} |\omega_k\rangle = \lambda_k |\omega_k\rangle, \ \langle \widetilde{\omega}_k | \mathcal{R} = \lambda_k \langle \widetilde{\omega}_k | \tag{A.4}$$

and

$$\langle \widetilde{\omega}_k | \omega_l \rangle = \delta_{kl}, \ \sum_k |\omega_k\rangle \langle \widetilde{\omega}_k | = 1 \tag{A.5}$$

Since $tr(\mathcal{R}) = N$, there are $N$ eigenvalues $\lambda_k$ equal to 1 that we will label with $k = 1, 2, \cdots, N$ and all the others are equal to zero. Finally, the definition of $\mathcal{R}$ and the constraints (A.1) immediately give:

$$\mathcal{R} |\varphi_n\rangle = |\varphi_n\rangle, \ \langle \psi_n | \mathcal{R} = \langle \psi_n | \tag{A.6}$$

It is thus possible to choose the orbitals of the two Slater determinants as the biorthogonal right and left eigenvectors of $\mathcal{R}$ associated to the unit eigenvalue: $|\omega_k\rangle = |\varphi_k\rangle$ and $\langle \widetilde{\omega}_k | = \langle \psi_k |$ for $k = 1, 2, \cdots, N$. In addition, the anticommutation rule $\{\hat{a}_{\widetilde{\omega}_k}, \hat{a}_{\omega_l}^+\} = \langle \widetilde{\omega}_k | \omega_l \rangle \hat{1} = \delta_{kl} \hat{1}$ between Fermi creation and annihilation operators directly imposes the identity:

$$\hat{a}_{\widetilde{\omega}_k} |\varphi\rangle = 0, \ \text{for any } k > N. \tag{A.7}$$

Conversely, the action of any one-body operator $\hat{A} = \sum_{i,j} A_{ij} \hat{a}_i^+ \hat{a}_j$ on the Slater determinant $|\varphi\rangle$ can be worked out through a decomposition of the single-particle states in the non-orthonormal eigenbasis of $\mathcal{R}$ according to the closure relation (A.5):

$$\hat{A}|\varphi\rangle = \sum_{k,l} \langle \widetilde{\omega}_k | A | \omega_l \rangle \, \hat{a}_{\omega_k}^+ \hat{a}_{\widetilde{\omega}_l} |\varphi\rangle \tag{A.8}$$

With the help of (A.7), only values of the integer $l$ ranging from 1 to $N$ give a non-zero contribution, and for $k \leq N$ the single-particle wave-function $|\omega_k\rangle = |\varphi_k\rangle$ belongs to the Fermi sea $\varphi$, so that $\hat{a}_{\omega_k}^+ \hat{a}_{\widetilde{\omega}_l} |\varphi\rangle = \delta_{kl} |\varphi\rangle$. One is finally left with:

$$\hat{A}|\varphi\rangle = \left[ tr(A\mathcal{R}) + \hat{A}_{0,1} \right] |\varphi\rangle \tag{A.9}$$

where $\hat{A}_{0,1}$ is the observable component that couples eigenspaces of $\mathcal{R}$ associated to different eigenvalues:

$$\hat{A}_{0,1} = \sum_{k>N, l\leq N} \langle \widetilde{\omega}_k | A | \omega_l \rangle \, \hat{a}_{\omega_k}^+ \hat{a}_{\widetilde{\omega}_l} = \sum_{i,j} \left[ (1-\mathcal{R}) A \mathcal{R} \right]_{ij} \hat{a}_i^+ \hat{a}_j, \tag{A.10}$$

By using this relation and $\left[\hat{A}, \hat{B}\right] = \sum_{i,j} [A,B]_{ij} \hat{a}_i^+ \hat{a}_j$ for one-body operators $\hat{A}, \hat{B}$, one can immediately check that a general two-body Hamiltonian (1) transforms the Slater determinant $|\varphi\rangle$ into:

$$\hat{H}|\varphi\rangle = \left[ \mathcal{H} + \hat{h}_{0,1} - \sum_{s\geq 1} \omega_s \hat{A}_{s,0,1}^2 \right] |\varphi\rangle. \tag{A.11}$$

$\mathcal{H}(\mathcal{R}) = \langle \psi | \hat{H} | \varphi \rangle = tr(A_o \mathcal{R}) - \sum_{s\geq 1} \omega_s \left[ tr^2(A_s \mathcal{R}) + tr(A_s (1-\mathcal{R}) A_s \mathcal{R}) \right]$ is the Hamiltonian $\hat{H}$ matrix element. $\hat{h}$ is a self-consistent one-body Hamiltonian defined by the matrix

$$h(\mathcal{R}) = A_o - \sum_{s\geq 1} \omega_s \left( A_s^2 + 2tr(A_s \mathcal{R}) A_s - 2 A_s \mathcal{R} A_s \right), \tag{A.12}$$

so that $h_{ij}(\mathcal{R}) = \dfrac{\partial \mathcal{H}}{\partial \mathcal{R}_{ji}}$ which exactly characterizes the Hartree-Fock Hamiltonian. Up to order $d\tau$, the imaginary-time infinitesimal propagation of the Slater determinant $|\varphi\rangle$ then also reads as

$$\exp(-d\tau \hat{H})|\varphi\rangle = \exp(-d\tau \mathcal{H}) \exp\left( -d\tau \left( \hat{h}_{0,1} - \sum_s \omega_s \hat{A}_{s,0,1}^2 \right) \right) |\varphi\rangle \tag{A.13}$$

This dynamics can be linearized with the Hubbard-Stratonovich transformation [19-21] allowing us to interpret each evolution under a quadratic form of one-body operators as the ensemble average (denoted $E(\cdots)$) over one-body evolutions in fluctuating auxiliary fields:

$$\exp\left( -d\tau \left( \hat{h}_{0,1} - \sum_s \omega_s \hat{A}_{s,0,1}^2 \right) \right) = E\left[ \exp\left( -d\tau \hat{h}_{0,1} + \sum_s dW_s \sqrt{2\omega_s} \, \hat{A}_{s,0,1} \right) \right] \tag{A.14}$$

where the $W_s$ are independent Wiener processes in the Itô stochastic calculus:

$$E(dW_s) = 0 , \quad dW_s dW_{s'} = d\tau \, \delta_{ss'} \tag{A.15}$$

The one-body propagators, obtained after application of the stochastic decoupling (A.14), now transform Slater determinants into new ones. Indeed, for any one-body operator $\hat{A}$, $\exp(\hat{A})|\varphi\rangle = |\varphi'\rangle$ where $|\varphi'\rangle$ is the Slater determinant with orbitals $|\varphi'_n\rangle = \exp(A)|\varphi_n\rangle$. Finally, to the first order in the imaginary-time step $d\tau$, the infinitesimal dynamics (A.13) can be reformulated as the weighted average of stochastic Slater determinants:

$$\exp(-d\tau \, \hat{H})|\varphi\rangle = E[\exp(-d\tau \mathcal{H}) \, |\varphi + d\varphi\rangle] \tag{A.16}$$

with the following variation of the Hartree-Fock orbitals:

$$|d\varphi_n\rangle = (1 - \mathcal{R})\left(-d\tau \, h(\mathcal{R}) + \sum_{s \geq 1} dW_s \sqrt{2\omega_s} \, A_s\right)|\varphi_n\rangle \tag{A.17}$$

Here, we have used the idempotency of $\mathcal{R}$ that implies $\mathcal{R}(1 - \mathcal{R}) = 0$. Note that he evolution process (A.17) preserves the biorthogonality constraints (A.1). Therefore, the propagation scheme (A.16) can be iterated for an arbitrary number of imaginary-time steps $d\tau$ and one is naturally led to the stochastic mean-field approach (3-5).

**Figure 1.** Estimate (7) of the energy as a function of the imaginary-time $\tau$ for the two-dimensionnal Hubbard model with a hole doping $\delta = 0.125$ from half-filling and an interaction parameter $U = 4t$. The trial Slater determinant for stochastic Hartree-Fock calculations (SHF) is indicated in parentheses. The results have been averaged over several hundred of runs of 100 trajectories. When not shown, statistical error bars are smaller than the symbol size. The black line gives on the left the exact ground-state energy [32] and on the right the constrained-path Monte-Carlo result [23].

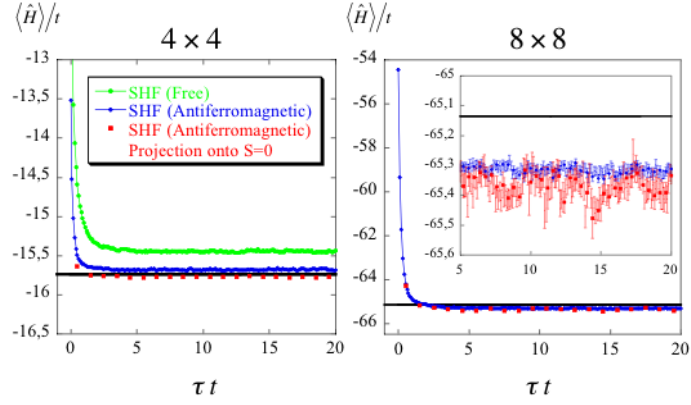

**Figure 2.** Extrapolated estimate (8) of the real space spin-spin correlation function for the $4 \times 4$ Hubbard model. $\delta$ is the hole doping and $t'$ denotes a next-to-nearest hopping. An antiferromagnetic mean-field solution was used as trial wave-function in all cases. Stochastic paths have been projected onto the spin-singlet sector. We averaged quantum Monte-Carlo results at the imaginary-time $\tau = 20/t$ with 100 trajectories over several hundred of runs. Statistical error bars are smaller than the size of the points.

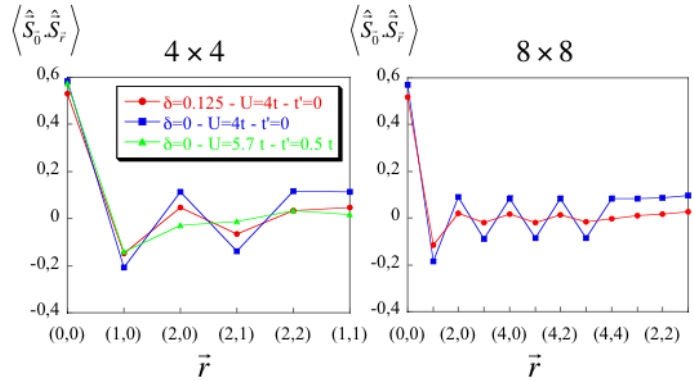

**Figure 3.** Stochastic Hartree-Fock calculations of the ground-state energy of a unitary Fermi gas with $N_\uparrow = N_\downarrow = N/2$ atoms in each spin state. $\eta(\tau)$ is the ratio between the mean-energy of at "time" $\tau$ and the ground-state energy of the non-interacting gas on the lattice. The trial state is a spin-singlet Slater determinant for the free gas. We show the average result over many runs of 100 paths. We use a $6\times 6\times 6$ lattice, except for $N \geq 40$ where the calculations were performed on a $8\times 8\times 8$ grid. The imaginary-time $\tau$ is expressed in units of $ml^2/\hbar^2$, where $l$ is the lattice spacing.

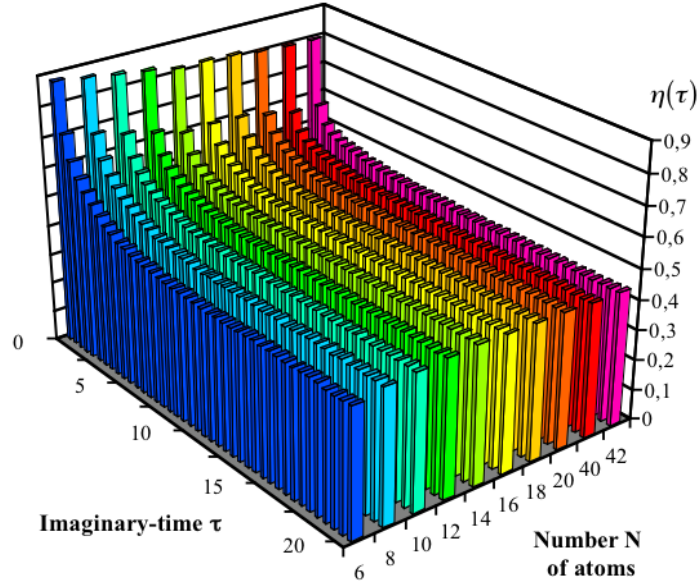

**Table 1:** Numerical values of the ratio $\eta = E_{g,\infty}/E_{g,0}$ for $N$ interacting fermions in the unitarity limit. $E_{g,a}$ is the ground-state energy corresponding to a physical scattering length $a$.

| Number $N$ of atoms | Ratio $\eta$ |
| --- | --- |
| 6 | 0.42551 |
| 8 | 0.44382 |
| 10 | 0.45316 |
| 12 | 0.45717 |
| 14 | 0.46012 |
| 16 | 0.45473 |
| 18 | 0.45273 |
| 40 | 0.45211 |
| 42 | 0.446 |